\documentstyle[12pt,preprint]{aastex}
\input epsf

\begin{document}

\title{The error budget of the Dark Flow measurement.}

\author{F. Atrio-Barandela\altaffilmark{1},
A. Kashlinsky\altaffilmark{2},
H. Ebeling\altaffilmark{3},
D. Kocevski\altaffilmark{4}
A. Edge\altaffilmark{5}}
\altaffiltext{1}{F{\'\i}sica Te\'orica, Universidad de Salamanca,
37008 Salamanca, Spain; atrio@usal.es}
\altaffiltext{2}{SSAI and Observational Cosmology Laboratory, Code
665, Goddard Space Flight Center, Greenbelt MD 20771}
\altaffiltext{3}{Institute for Astronomy, University of Hawaii, 2680
Woodlawn Drive, Honolulu, HI 96822}
\altaffiltext{4}{Department of Physics, University of California at Davis,
1 Shields Avenue, Davis, CA 95616}
\altaffiltext{4}{Department of Physics, University of Durham, South Road,
Durham DH1 3LE, UK}


\begin{abstract}
We analyze the uncertainties and possible systematics associated
with the ``Dark Flow" measurements using the cumulative
Sunyaev-Zeldovich effect combined with all-sky catalogs of
clusters of galaxies. Filtering of all-sky cosmic microwave
background (CMB) maps is required to remove the intrinsic
cosmological signal down to the limit imposed by cosmic variance.
Contributions to the errors come from the remaining cosmological
signal, that integrates down with the number of clusters, and the
instrumental noise, that scales with the number of pixels; the
latter decreases with integration time and is subdominant for the
Wilkinson Microwave Anisotropy Probe 5-year data. It is proven
both analytically and numerically that the errors for 5-year WMAP
data are $\simeq 15\sqrt{3/N_{\rm clusters}} \mu$K per dipole
component. The relevant components of the bulk flow velocity are
measured with a high statistical significance of up to $\ga
3-3.5\sigma$ for the brighter cluster samples. We discuss different
methods to compute error bars and demonstrate that they have
biases that would over predict the errors, as is the case in a
recent reanalysis of our earlier results. If the signal is caused
by systematic effects present in the data, such systematics must
have a dipole pattern, correlate with cluster X-ray
luminosity and be present only at cluster positions. Only
contributions from the Sunyaev-Zeldovich effect could provide such
contaminants via several potential effects. We discuss such
candidates apart from the bulk-motion of the cluster samples and
demonstrate that their contributions to our measurements are
negligible. Application of our methods and database to the
upcoming PLANCK maps, with their large frequency coverage, and in
particular, the 217GHz channel, will eliminate any such
contributions and determine better the amplitude, coherence and
scale of the flow.
\end{abstract}


\section{Introduction.}

Peculiar velocities are deviations from the uniform expansion of
the Universe. In the gravitational instability model, they are
generated by the inhomogeneities in the matter distribution. Most
determinations of the peculiar velocities are based on surveys of
individual galaxies. Early measurements by Rubin et al (1976)
found peculiar flows of amplitudes $\sim$700 km/sec. Using the
``Fundamental Plane" (FP) relation (Dressler et al 1987,
Djorgovski \& Davis 1987) suggested that elliptical galaxies
within $\sim 60h^{-1}$Mpc were streaming at $\sim 600$ km/sec with
respect to the CMB (Lynden-Bell 1988). Mathewson et al (1992) used
the Tully-Fisher (1977 - TF) relation and found that this flow of
amplitude 600 km/sec does not converge until scales much larger
than $\sim 60 h^{-1}$ Mpc, a result that was in agreement with a
later analysis by Willick (1999). Using the brightest galaxy as a
distance indicator for a sample of 119 rich clusters, Lauer \&
Postman (1994 - LP) measured a bulk flow of $\sim$700 km/sec on a
scale of $\sim$150$h^{-1}$Mpc, but a re-analysis of these data by
Hudson \& Ebeling (1997) taking into account the correlation
between the luminosity of brightest galaxy and that of its host
cluster found a reduced bulk flow pointing in a different
direction. Using the FP relation for early-type galaxies in 56
clusters Hudson et al (1999) found a similar bulk flow as LP and
on a comparable scale, but in a different direction. A sample of
24 SNIa showed no evidence of significant bulk flows out to $\sim
100 h^{-1}$ Mpc (Riess et al 1997), and a similar conclusion was
reached with a TF-based study of spiral galaxies by Courteau et al
(2000). Kocevski \& Ebeling (2006) analyzed the contribution to
the peculiar velocity of the Local Group due to structures beyond
the Great Attractor and found that the dipole anisotropy of the
all-sky, X-ray-selected cluster sample compiled there suggested
that most of the flow was due to overdensities at $\ga
150h^{-1}$Mpc. Watkins et al (2009) developed a method to suppress
the sampling noise in the various galaxy surveys and showed that
all the data (except for the LP sample) agreed with substantial
motion on a scale of $\simeq 50-100 h^{-1}$Mpc. In a follow-up
study Feldman et al (2009) estimated the source of the flow to be
at an effective distance larger than 200$h^{-1}$Mpc; they
suggested that the absence of shear is consistent with the
attractor being at infinity, as proposed in Kashlinsky et al
(2008, hereafter KABKE1).

Cosmic Microwave Background (CMB) temperature fluctuations in the
direction of clusters of galaxies provide an alternative method to
measure peculiar velocities. The scattering of the microwave
photons by the hot X-ray emitting gas inside clusters induces
secondary anisotropies (Sunyaev \& Zeldovich 1970, 1972) that are
redshift independent and, if the noise is isolated, can be used to
probe the velocity field to much higher redshifts than with
galaxies. The Sunyaev-Zeldovich (SZ) effect has two components:
thermal (tSZ), due to thermal motions of electrons in the
potential wells of clusters and kinematic (kSZ) due to motion of
the cluster as a whole with respect to the isotropic CMB rest
frame (see review by Birkinshaw 1999). However, such measurements
for individual clusters are dominated by large errors.  On all-sky
CMB maps, the bulk flow motion of clusters of galaxies can be
obtained by using large all-sky cluster samples and evaluating the
CMB dipole at cluster locations (Kashlinsky \& Atrio-Barandela,
2000 - hereafter KAB). We have applied the KAB method using the
largest - at that time - sample of galaxy clusters in conjunction
with the 3-yr WMAP data and have uncovered a large scale flow of
amplitude $600-1000$km/s extending to at least $\simeq 300h^{-1}$
Mpc (Kashlinsky et al 2008, 2009a - KABKE1,2). That analysis has
now been extended to a still larger and deeper sample of over
1,000 clusters and 5-yr WMAP data (Kashlinsky et al 2009b -
hereafter KAEEK). The KAEEK analysis confirms the KABKE1,2 results
and shows that the flow remains coherent and extends to at least
twice the distance probed in KABKE1,2. A larger cluster sample
enabled KAEEK to bin the signal by cluster X-ray luminosity
($L_X$). The dipoles evaluated for binned subsamples increase
systematically with increasing $L_X$-threshold, as expected if the
signal is produced from the kSZ effect by all clusters
participating in the same motion, a correlation that would not
exist if the signal was produced by a rare excursion from noise or
primary CMB. 

Upcoming data from both the long integration WMAP data and the
PLANCK mission will bring more accurate CMB maps. PLANCK data will
be particularly important because of its wider frequency coverage,
finer angular resolution and lower instrument noise. It is
imperative to identify the prospects and limitations of the
applications of the current KAB methodology to these future
datasets. Also, alternative methods can test the
existence of a large scale flow (see Itoh, Yahada \& Takata, 2009;
Zhang, 2010).

In this paper we present a detailed analysis of the uncertainties
affecting the measured bulk flow providing the necessary details
to support the KAEEK results. In Sec. 2 we briefly summarize our
method followed by a summary of data and dipole analysis. In Sec.
3 we present a theoretical derivation of the error bars, showing
when they become dominated by cosmic variance of the cosmological
CMB residual that remains in the maps after filtering. It is
demonstrated that for the CMB sky of {\it our} Universe and an
isotropic all-sky cluster catalog the errors in the KABKE/KAEEK
dipole measurements are $\simeq 15\sqrt{3/N_{\rm cl}}\mu$K per
dipole component. {\it Given the filtering scheme adopted in our
studies}, these errors cannot be reduced much in CMB data with
lower instrument noise. Rather the strategy of further increasing
the signal-to-noise in the measured dipole value must be through
increasing the number of clusters. Particularly important would be
to increase the number of observed clusters at the bright end of
the cluster luminosity function, where the much larger cluster
optical depth, $\tau$, compensates for the decrease in the
abundance of such clusters (KAEEK). In Sec 4 we discuss different
methods to estimate error bars describing their various biases.
Sec. 5 addresses the overall statistical significance of the
measurement. Given that the measured dipole increases with the
X-ray luminosity threshold, the signal found in KABKE1,2 and KAEEK
cannot arise in primary CMB, but tSZ dipole contributions can
potentially provide confusion to the measurement. Section 6
discusses such possible systematic effects due to tSZ and we show
there that all are of negligible amplitude and none could have
generated the measured signal. Finally, in Sec. 7 we present our
conclussions.

\section{Methodology and Analysis.}

\subsection{KA-B method}

If a cluster at angular position $\vec{y}$ has the line-of-sight
velocity $v$ with respect to the CMB, the SZ CMB fluctuation at
frequency $\nu$ at this position will be $\delta_\nu(\vec
y)=\delta_{\rm tSZ}(\vec y)G(\nu)+ \delta_{\rm kSZ}(\vec
y)H(\nu)$, with $ \delta_{\rm tSZ}$=$\tau T_{\rm X}/T_{\rm e,ann}$
and $\delta_{\rm kSZ}$=$-\tau (v_p/c)\cos\theta$, being $\theta$
the angle between the cluster peculiar velocity $\vec{v}_p$ and
the line of sight. Here $G(\nu)\simeq-1.85$ to $-1.35$ and
$H(\nu)= 1$ if the thermodynamic CMB temperature is measured over
the range of frequencies probed by the WMAP data, $\tau$ is the
projected optical depth due to Compton scattering, $T_{\rm X}$ is
the cluster electron temperature and $k_{\rm B}T_{\rm e,ann}$=511
KeV. When averaged over many isotropically distributed clusters
moving with a significant bulk flow with respect to the CMB, the
kinematic term generates a dipole contribution that could
dominate, enabling a measurement the bulk motion $V_{\rm bulk}$ of
the cluster sample. Thus, KAB suggested measuring the dipole
component of $\delta_\nu(\vec y)$ at $N_{\rm cl}$ cluster
locations on the CMB sky.

We denote by $(a_0,a_{1x},a_{1y},a_{1z})$ the monopole and three
dipole components evaluated over some locations in the sky and
follow the same conventions as in KABKE1,2 and KAEEK:
$a_0=\langle\Delta T\rangle$ and $a_{1i}=\langle\Delta T
n_i\rangle$ with $i=(x,y,z)$ and
$n_{x,y,z}=(\sin\theta\cos\phi,\sin\theta\sin\phi,\cos\theta)$ are
the direction cosines of a vector with angular coordinates
$(\theta,\phi)$. Brackets represent averages taken over the
cluster population of our catalog. The definition of monopole and
dipole above follows the convention used in the Healpix {\bf
remove\_dipole} routine.  The dipole power is defined as $C_1=
\sum_{m=-1}^{m=1} |a_{1m}|^2$, where $a_{1m}$ are the three dipole
components. With our normalization, $C_{1,{\rm kin}}$ is such that
a coherent motion at velocity $V_{\rm bulk}$ would lead to
$C_{1,{\rm kin}}= T_{\rm CMB}^2 \langle \tau \rangle^2 V_{\rm
bulk}^2/c^2$, where $T_{\rm CMB} =2.725$K is the present-day CMB
temperature. For reference, $\sqrt{C_{1,{\rm kin}}}\simeq 1
(\langle \tau \rangle/10^{-3}) (V_{\rm bulk}/100{\rm km/sec}) \;
\mu$K.

When the dipole is computed at the position of $N_{\rm cl}$
clusters, it will have contributions from 1) the instrument noise,
2) the tSZ  component, 3) the primary cosmological CMB fluctuation
component from the last-scattering surface, and 4) the various
foreground contributions at WMAP frequencies. The latter can be
significant at the K and Ka WMAP channels, so we restricted our
analyses to the WMAP Channels Q, V and W which have negligible
foreground contributions.

For $N_{\rm cl}\gg 1$ the dipole of the observed $\delta_\nu$
becomes
\begin{equation}
a_{1m} \simeq a_{1m}^{\rm kSZ} +a_{1m}^{\rm tSZ} + a_{1m}^{\rm
CMB} + \frac{\sigma_{\rm noise}}{\sqrt{N_{\rm cl}}}
\label{eq:dipole}
\end{equation}
Prior to any analysis, the CMB dipole due to our motion with
respect to the isotropic CMB frame is removed from the data. The
KSZ effect measures velocities with respect to the CMB frame which
also is taken to be the frame of the universal expansion. This
does not change when all-sky dipole or any other $\ell$-pole
moments are subtracted in the all-sky maps. This dipole
subtraction removes our peculiar velocity, $v_{\rm local}$,
contributions down to $O[(v_{\rm local}/c)^2]$ contributions to
the quadrupole. To check that the latter does not contribute to
the measurement, we also ran the pipeline subtracting the all-sky
quadrupole from the original maps and detected only negligible
differences in the final results. As shown in KAB, in this way the
kSZ term can be isolated in eq.~\ref{eq:dipole}.

\subsection{Map preparation and analysis}

The process that enabled us to isolate the kSZ term is described
in detail in KABKE1,2. Briefly:

(A) An all-sky catalog of X-ray selected galaxy clusters was
constructed using available X-ray data extending to $z\simeq 0.3$.

(B) As indicated, we only used WMAP Q, V and W bands, where the
foreground contamination is smallest.  We applied the 3- and 5-yr
version of the Kp0 mask to remove those pixels where galactic or
point source contributions dominate. Next, to prevent any power
leakage from the dipole generated by our peculiar velocity, it was
removed from the pixels that survived the mask. Furthermore, KAEEK
explicitly removed dipole {\it and} quadrupole from the original
maps and demonstrated that the quadrupole did not contribute to
the results. This removes $v_{\rm local}$ down to $O[(v_{\rm
local}/c)^3]$ contribution to the octupole.

(C) The cosmological CMB component was removed from the WMAP data
using a Wiener-type filter, constructed using the $\Lambda$CDM
model that best fit the data.  It was constructed in order to
minimize the difference $\langle (\delta T - {\rm
noise}))^2\rangle$. Next, filtered maps were constructed using all
multipoles with $\ell\ge 4$ and keeping the same phases as in the
original maps. Modes with $\ell\le 3$  were not included to avoid
any possible contributions that could be introduced by the
alignment of those low order multipoles and also because those
modes would potentially be the most affected by any hypothetical
power leakage.

(D) The Wiener filter is constructed (and is
different) for each DA channel because the beam and the noise are
different. This prevents inconsistencies and systematic errors
that could have been generated if a common filter was applied to
the eight channels of different noise and resolution.

(E) In the filtered maps, the monopole and dipole are computed
exclusively at the cluster positions, using Healpix {\bf
remove\_dipole} routine ascribing to each cluster a given circular
aperture. Due to the variations of the Galactic absorbing column
density and ROSAT observing strategy, cluster selection function
and X-ray properties may vary across the sky introducing possible
systematics. In KABKE1,2 we used the measured X-ray extent of each
cluster, $\theta_X$ and computed the dipole for different
apertures, in multiples of $\theta_X$ and, to avoid being
dominated by a few very extended nearby clusters like Coma, we
introduced a cut so the final extent of any cluster was always
smaller than 30'. There we computed core radii directly from the
data and from an $L_X-r_c$ relation. Analyses using both sets gave
consistent results, consistent with the X-ray systematic effects
not affecting our results significantly. More important,
variations in the final aperture were already tiny in the KABKE1,2
analysis and KAEEK used altogether a fixed aperture were the mean
monopole vanishes. The KAEEK results are consistent with the
previous (KABKE1,2) measurements. Fixing the same aperture for all
clusters simplifies the statistical analysis and this is the
approach taken in this article.

(F) We compute the monopole and dipole for different angular
apertures. At small apertures ($\sim 10^\prime$), clusters show a
clear tSZ decrement, but the amplitude of the signal falls off
with increasing angular aperture. The final dipole is computed at
the aperture where the mean monopole of the clusters vanishes.
This ensures that the TSZ contribution to the measured dipole is
negligible and does not confuse the KSZ component.

(G) Our final result is a dipole measured in units of
thermodynamic CMB temperature. To translate the three measured
dipoles into three velocity components, we need to determine the
average cluster optical depth to the CMB photons,
$\langle\tau\rangle$, on the filtered maps. Since filtering
reduces the intrinsic CMB contribution, it also modifies its
optical depth, $\tau$. In KABKE1,2 we introduced a calibration
factor $C_{1,100}$ that gave the kSZ dipole in $\mu$K of a bulk
motion of amplitude $V_{\rm bulk}=100km/s$. The calibration factor
depends both on the filter and on the cluster profile. In KABKE1,2
and KAEEK it was estimated using a $\beta$ model and the angular
X-ray extent of the cluster.

We defer to Sec.~5 a discussion on the statistical significance of
our measurements. We emphasize that in the filtered maps we
measure monopole and dipole simultaneously. The monopole is
dominated by the tSZ component and its amplitude sets an upper
limit on $a_{1m}^{tSZ}$ (see eq.~\ref{eq:dipole}), the tSZ dipole
due to an inhomogeneous cluster distribution on the sky. We found
a dipole at cluster positions with a high confidence level and we
obtained this dipole at the (fixed) cluster aperture {\it when the
tSZ monopole component was zero}. Since the tSZ component from the
clusters vanishes, only a contribution from the kSZ component, due
to large-scale bulk motion of the cluster sample, remains.

The main current uncertainty in our method is the calibration,
currently parameterized with the $C_{1,100}$ quantity, which
generally is a matrix.
At present, we do not have enough information on the tSZ profile
of the clusters in our catalog to increase the accuracy of our
calibration. Sec. 8 of KABKE2 discusses the issues and points out
that we may be {\it overestimating} the velocity amplitude in the
current cluster catalog by $\sim 20-30 \%$. However, this question
is significant only insofar as the precise amplitude of the flow
velocity in km/sec is concerned. PLANCK, with its large frequency
coverage, will allow the measurement of individual profiles for an
important fraction of clusters in our catalog and should enable us
determine the calibration coefficients more accurately.

\section{Noise and Intrinsic CMB Residual Contributions.}

The KAB method to measure bulk flows using clusters of galaxies as
tracers of the velocity field requires the intrinsic CMB component
to be removed from the data. To this end we have designed a filter
which minimizes $\langle (\delta T-{\rm noise})^2\rangle$. As we
show below this filter removes the primary CMB anisotropies down
to the fundamental limit imposed by the cosmic variance. In
Fourier space this filter is expressed as
\begin{equation}
F_\ell=\frac{|d_\ell|^2-C_\ell^{th}B_\ell^2}{|d_\ell|^2}
\end{equation}
where $|d_\ell|^2=(2\ell+1)^{-1}\sum_m |a_{\ell m}|^2$ is the
power measured in each Differencing Assembly (DA) corrected for
the mask sky area, and $C_\ell^{th}B_\ell^2$ is the power spectrum
of the theoretical model that best fits the data, convolved with
the antenna beam $B_\ell$ of each DA. Although this filter removes
much of the intrinsic primary CMB contributions, it leaves a
residual CMB component since the theoretical model does not
reproduce perfectly the data measured at our location. This
residual will be common to all frequencies and, since it is
correlated between the various DA's, it limits the accuracy down
to which the primary CMB can be removed in the KAB method.

Because of the cosmic variance, the power of the CMB sky at our
location $C_\ell^{LOC}$ differs from the theoretical model
$C_{\ell}^{th}$ and so a residual CMB signal from primary
anisotropies is left in the filtered maps. To estimate the
contribution of noise and the CMB residual to the total power on
these maps, let $\delta T(\hat{n}) = \sum F_\ell a_{\ell m}Y_{\ell
m}(\hat{n})$ be the temperature anisotropy of the filtered maps
expanded in spherical harmonics $Y_{\ell m}$. The variance of any
filtered map is:
\begin{equation}
\sigma_{fil}^2=\frac{1}{4\pi}\sum (2\ell+1)F_\ell^2|d_\ell|^2=
\frac{1}{4\pi}\sum
(2\ell+1)\frac{(|d_\ell|^2-C_\ell^{th}B_\ell^2)^2}{|d_\ell|^2} .
\label{sigmafil}
\end{equation}
As indicated, $\delta T(\hat{n})$ contains the cosmological CMB
signal and noise, $|d_\ell|^2=C_\ell^{LOC}B_\ell^2+N_\ell$. The
power spectrum at our location differs from the underlying power
spectrum by a random variable of zero mean and (cosmic) variance
$\Delta_\ell=(\ell+\frac{1}{2})C^{th}_\ell/f_{sky}$, where
$f_{sky}$ is the fraction of the sky covered by the data (Abbot \&
Wise 1984). Then, due to cosmic variance,
$C_\ell^{LOC}=C_{\ell}^{th}\pm\Delta_\ell^{1/2}$. The above limits
on $C_\ell$ bound the range of $\sigma_{fil}$,
eq.~(\ref{sigmafil}), to:
\begin{equation}
\sigma_{fil}^2=\frac{1}{4\pi}\sum (2\ell+1)\left[\frac{\Delta_\ell^2}
{C_\ell^{th}+\Delta_\ell+N_\ell}+\frac{N_\ell^2}{C_\ell^{th}+\Delta_\ell+N_\ell}\right]=
\sigma_{CV,fil}^2+\sigma_{N,fil}^2(t_{\rm obs}) \label{sigmafil2}
\end{equation}
In this last expression, the variance of the filtered map depends
on two components: the residual CMB left due to cosmic variance
$\sigma_{CV,fil}$ and the noise $\sigma_{N,fil}$, that is
not removed by the filter. The latter component integrates down with
increasing observing time $t_{\rm obs}$ as
$t_{\rm obs}^{-1/2}$ and becomes progressively less important in
WMAP data with longer integration time.

We denote by $\sigma_q^2\equiv\frac{1}{4\pi}(2q+1)
(\Delta^2_q+N^2_q)(C_q^{th}+\Delta_q+N_q)^{-1}$ and let
$\sigma^2(\ell)=\sum_{q=4}^\ell\sigma_q^2$ be the cumulative
variance of the residual map. With these definitions, the total
variance of the filtered map is $\sigma_{fil}^2=
\sigma_{fil}^2(\ell_{max})$. For Healpix maps with $N_{side}=512$
the maximal multipole is $\ell_{max}=1024$ (Gorski et al 2005). In
Figure~\ref{fig1} we plot this cumulative contribution of each
multipole $\ell$, $\sigma_{fil}(\ell)$, to the total rms of the
map. The solid lines represent the mean and rms
$\sigma_{fil}(\ell)$ of filtered maps of 4,000 realizations of the
Q1 DA; the shaded area represents the dispersion of those
realizations, the dot-dashed line is the same quantity but for the
filtered Q1 WMAP 5-year data. The lower dashed lines represent
$\sigma_{CV,fil}$, the residual CMB component, and upper dashed
line, the total variance of the map [eq~(\ref{sigmafil2})]. The
dot-dashed line also contains any contributions from foreground
emissions; the fact that it lies so close to the to the region
expected from simulating CMB sky implies that foreground emission
contributions to $\sigma_{fil}$ are small. Figure~\ref{fig1}
clearly shows that for multipoles below $\ell\sim 200$ the
cumulative variance of the 5-year WMAP maps $\sigma^2(\ell)$ is
dominated by the residual primary CMB signal from the cosmic
variance, even though the total variance of the filtered maps is
dominated by noise. For the Q1 WMAP channel, the mean variance of
our simulations was $\sigma_{fil}^2\sim 2000 (\mu$K)$^2$ out of
which $\sim 200 (\mu$K)$^2$ come from the residual primary CMB
signal.

Finally, Figure~\ref{fig1} indicates that our filter removes the
intrinsic CMB down to the fundamental limit imposed by cosmic
variance. In this sense the filter is close to optimal, since it
minimizes the errors contributed to our measurements by primary
CMB. In principle, one can define a more aggressive filter that,
together with the intrinsic CMB, also removes the noise leaving
only the SZ signal. But filtering is not a unitary operation and
does not preserve power. Such a filter would then remove an
important fraction of the SZ component and would probably reduce
the overall S/N. In general, a different filter would give
different dipole (measured in units of temperature) and would
require a different calibration. Discussion of filtering schemes
that maximize the S/N ratio and minimize the systematic error on
the calibration will be given elsewhere.

\section{Monopole and dipole uncertainties.}

Here we consider how the two components present in the filtered
maps, i.e. 1) residual primary CMB and 2) instrument noise,
contribute to the uncertainty in the measurement of bulk flows. In
KABKE1,2 we adopted two methods to estimate the uncertainties: (I)
evaluating monopole and dipole on the filtered maps outside
cluster locations and (II) using the same cluster template on
simulated maps. Both methods are different but complementary.
Errors estimated using method I include any contribution
originated by foreground residuals and CMB masking while in Method
II we account for the inhomogeneity of the cluster distribution on
the sky.

It is important to emphasize that the filtered maps have no {\it
intrinsic monopole or dipole} by construction. Since we measure
these two moments from a small fraction of the sky, our limited
sampling generates an error due to (random) distribution of these
quantities around their mean zero value. The sampling variances of
$\langle a_0\rangle$ and $\langle a_{1i}\rangle$ are $Var(\langle
a_0\rangle)=\langle a_0^2\rangle/N$, $Var(\langle
\sigma_{i}\rangle)=\langle a_{1i}^2\rangle/N$, where $N$ is the
number of independent data points. Direct computation shows that:
\begin{equation}
\sigma_0^2\equiv \langle a_0^2\rangle = \langle (\Delta
T)^2\rangle \qquad \sigma_i^2\equiv \langle a_i^2\rangle =
\frac{\langle (\Delta T)^2\rangle}{\langle n_i^2\rangle}, \quad
i=(x,y,z)
 \label{eq:sqrt3}
\end{equation}
In this expression, $n_i$ are the direction cosines of clusters.
If clusters were homogeneously distributed on the sky then
$\langle n_i^2\rangle=1/3$ and one should recover the dipole
errors of $\sigma_i=\sqrt{3}\sigma_0$. Thus the error on the
monopole serves as a consistency check in any such computation. 

\begin{figure}
\plotone{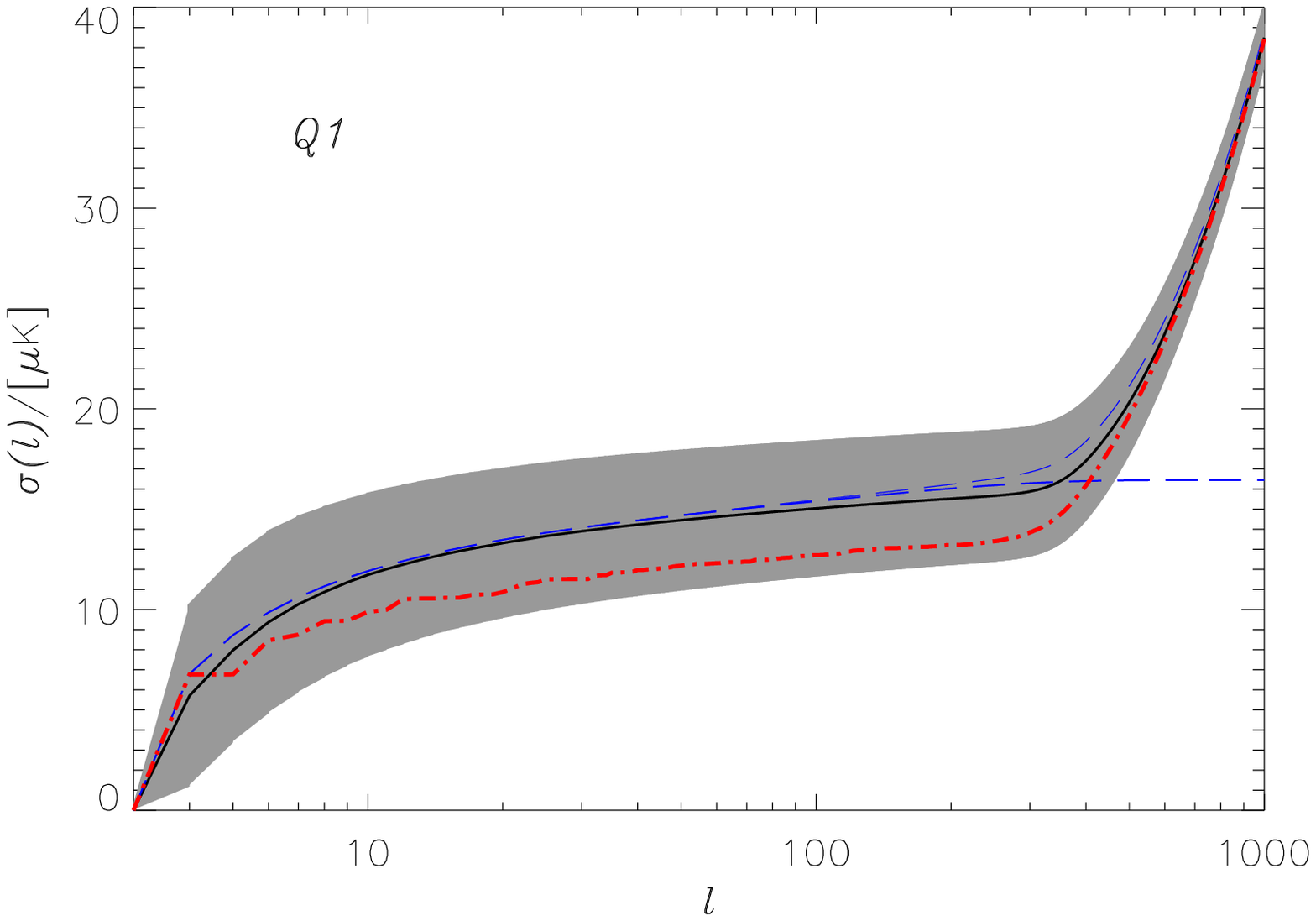} \caption{\small{ Cumulative rms deviation as
a function of multipole. Solid line and shaded area show mean and
rms of 4,000 simulated Q1 filtered maps. Dashed lines represent
the residual CMB component of the filtered maps due to cosmic
variance, computed using eq.~(\ref{sigmafil2}) and the residual
CMB plus the noise components. The dot-dashed line corresponds to
the actual Q1 band of WMAP 5-yr data. }}
 \label{fig1}
\end{figure}

Sec. 3 discussed the two components of the variance of the
filtered map. As before, $\sigma_{CV,fil}^2$ and
$\sigma_{N,fil}^2$ represent the contribution to the total
variance due to the residual CMB component and the noise,
respectively. When we estimate error bars by placing random
clusters on the real filtered maps outside clusters (Method I) or
the real clusters on simulated filtered data (Method II), $N_{cl}$
clusters occupy $N_{pix}$ in $N_{DA}$ Differencing Assemblies. As
the residual CMB signal is correlated from map to map, it will
decrease only as the number of clusters increases, but the noise
term will decrease much faster since it is uncorrelated from map
to map and pixel to pixel and integrates down with $N_{DA}$ and
the integration time. Then, the resulting sampling variance will
be
\begin{equation}
\sigma_{0}^2=\frac{\sigma_{CV,fil}^2}{N_{cl}}+
\frac{\sigma_{N,fil}^2(t_{\rm obs})}{N_{DA}N_{pix}},\qquad
\sigma_{i}^2=\frac{\sigma_{0}^2}{\langle n_i^2\rangle}=3\sigma_{0}^2
\label{sigma_mon}
\end{equation}
As expected from eq. \ref{eq:sqrt3},  for an homogeneous cluster
catalog the variance in each dipole component is three times
larger than on the monopole since three quantities are derived
from the same data set. From Figure~\ref{fig1} we obtain that
$\sigma_{CV,fil}\simeq 15\mu$K and $\sigma_{N,fil}\simeq 40\mu$K.
When clusters are not homogeneously distributed in the sky, the
basis of direction cosines is no longer orthogonal and error bars
need to be estimated numerically.

The results presented in Figure~\ref{fig1} together with
eq.~(\ref{sigma_mon}) indicate that Method II will give slightly
larger error bars. If monopole and dipole are evaluated at cluster
positions on {\it simulated} maps then $\sigma_{CV,fil}$ and
$\sigma_{N,fil}$ in eq.~(\ref{sigma_mon}) will be close to the
average CMB residual and noise of the simulated maps. As
Figure~\ref{fig1} indicates, they are larger than the filtered
data (shown by a dot-dashed line) corresponding to the CMB
realization representing our Universe. The latter, however, is the
only CMB sky relevant for the true error analysis in this
measurement. This fact was already noticed in KABKE2 where such
comparison was made and the errors were found to be 10-15\% larger
if using Method II.

\begin{figure}
\plotone{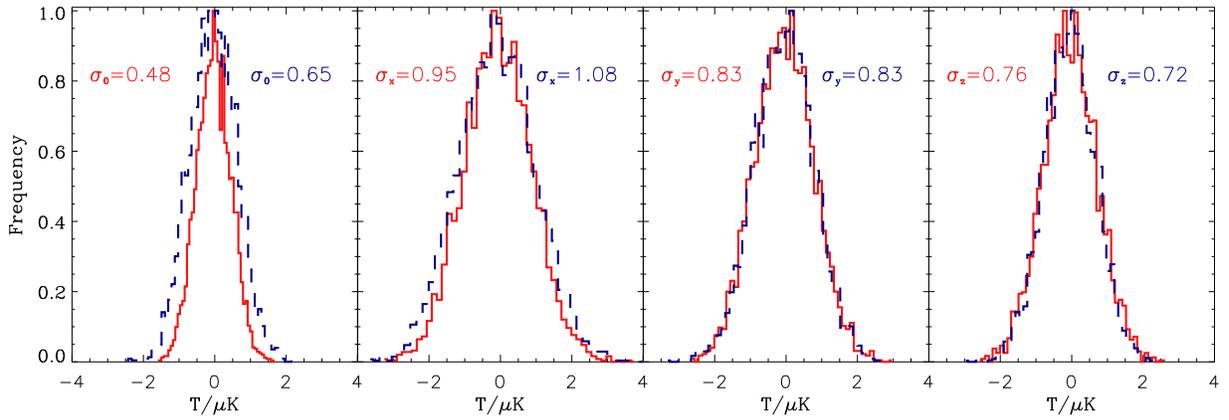} \caption{\small{ Histograms of the
distribution of monopoles and the three dipole components computed
using the filtered Q1 WMAP 5-yr map data. Solid, dashed lines
correspond to Method I and Method IIa  of 4,000 simulations (see
text), respectively. Also indicated is the rms dispersion
(in micro Kelvin)  for Method I (left) and Method IIa (right). } }
 \label{fig2}
\end{figure}

To avoid this bias, we introduce Method IIa: error bars are
computed from random realizations of the {\it power spectrum of
the filtered maps}. In Figure~\ref{fig2} we plot the histograms of
the monopole and dipole components of 4,000 simulations of 1000
clusters with constant angular size of 30$^\prime$ with both
Method I (random clusters located outside the mask on the real
data) and Method IIa (the cluster template is fixed and the sky is
simulated; the spectrum are gaussian realizations of the measured
power of the filtered maps). From left to right we display the
histogram of the monopole and $(x,y,z)$ components of the dipole.
The rms deviations, given to the left and right of each plot,
correspond to Method I and Method IIa, respectively. Solid lines
represent the histograms in Method I and dashed lines in Method
IIa. We find that, to good accuracy, the distribution of the
monopole and dipoles is Gaussian with zero mean. More importantly,
we see no systematic differences between both methods. Then,
neither foreground residuals nor cluster inhomogeneities have a
significant contribution to the estimated error bars. Instead the
errors are dominated by the sampling/cosmic variance when
measuring the monopole and dipole from a limited fraction of the
sky.

\begin{figure}
\plotone{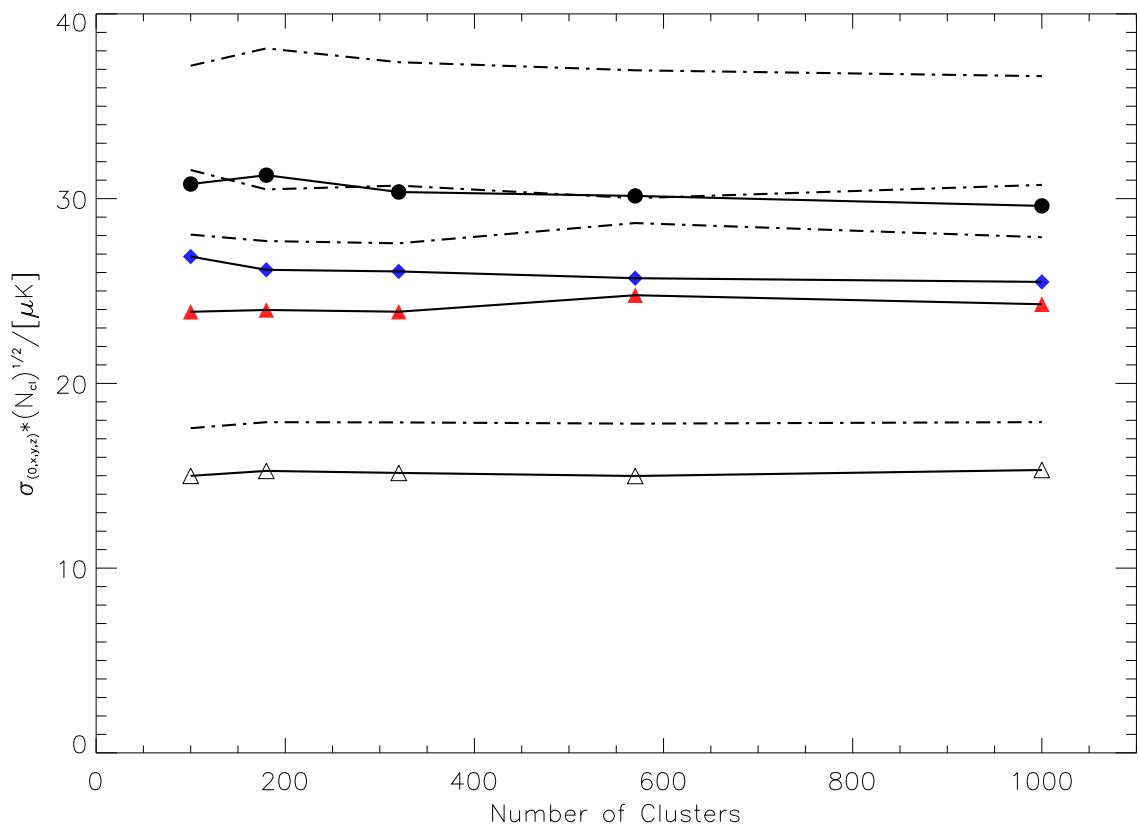}
\caption{\small{ (a) Rms deviation of the
monopole and three dipole components computed scaled by the
number of clusters. Open triangles, circles, diamonds (blue) and
solid triangles (red) correspond to the monopole and (x,y,z) components of the dipole.
Solid lines joint the symbols of clusters with 30' radius, while
dashed lines follow the same ordering than solid lines but correspond
to clusters with 20' radius.
}}
\label{fig3}
\end{figure}

To test the validity of eq.~(\ref{sigma_mon}) we carried out
another 4,000 simulations with different number of clusters:
$N_{\rm cl}=$100, 180, 320, 570 and 1000 in accordance with Table
1 of KAEEK. In Method I, we placed $N_{cl}$ clusters at random on
the sky. To be fully consistent with how cluster samples are
selected from the data, the smaller samples are subsets chosen
randomly from the full sample. In Figure~\ref{fig3} we plot the
rms deviation of the monopole (open triangles) and the three
dipole components. Filled circles, diamonds and solid triangles
correspond to the $(x,y,z)$ dipole components.  Solid lines
connect the results when clusters are assigned radii of
30$^\prime$, while dashed lines correspond to results with
20$^\prime$ clusters and follow the same ordering as the solid
lines. The figure shows that $\sigma_{(0,x,y,z)}\propto
N_{cl}^{1/2}$ with great accuracy. As expected, the errors are
larger when the cluster size is smaller because of the different
numbers of pixels entering the instrument noise contribution in
eq. (\ref{sigmafil2}). In Figure~\ref{fig3}, the differences
between the dipole components come from differences in sky
coverage. The $x$ and $y$ components, that are in the plane of the
Galaxy, are determined with progressively less accuracy since the
CMB data in the Galactic plane is dominated by foreground
emission. Still the difference from the uncertainty of the
$z$-component is small ($\la 10\%$) particularly for the better
measured $y$-component of the dipole.

\begin{figure}
\plotone{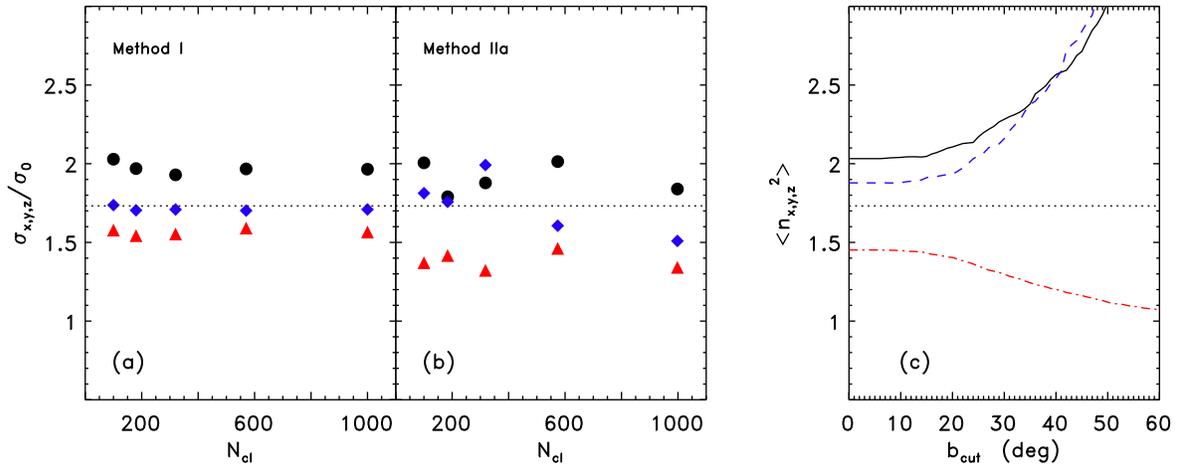} \caption{\small{ Dipole to monopole error
bar ratio. (a) (Black) circles, (blue) diamonds and (red) circles
correspond the the ratio of the (x,y,z) component of the dipole to
the monopole, respectively. Monopole and dipole were computed
using Method I. (b) Same as (a) but monopole and dipoles are
computed using Method IIa. (c) Ratio of the dipoles to monopole
error bars for our cluster catalog. The horizontal axis, $b_{cut}$
indicates that clusters with $|b|\le b_{cut}$ are excised from the
catalog. In all three plots, the dotted line represents the ratio
for a perfectly isotropic cluster catalog. }} \label{fig4}
\end{figure}

We can use eq.~(\ref{sigma_mon}) to estimate how accurately we
measure any dipole component compared to the monopole. In
Figure~\ref{fig4}a we plot $\sigma_{(x,y,z)}/\sigma_0$,
the ratio of the rms deviation of
dipole components to the rms of the monopole of the 4,000
simulations generated using Method I, as described above. Filled
circles, diamonds and triangles correspond to the ratio of the
$(x,y,z)$ rms deviation of the dipole to that of the monopole,
respectively. In Figure~\ref{fig4}b we plot the same magnitudes
for Method IIa. The dotted line represents
$\sigma_{(x,y,z)}/\sigma_0=\sqrt{3}$ that corresponds (in
eq~\ref{sigma_mon}) when clusters are homogeneously distributed in
the sky. Figure~\ref{fig4}a clearly indicates that when clusters
are chosen randomly on the sky, the error on the x-component is
larger than on y or z and the scaling with the number of clusters
was very close to $N_{cl}^{1/2}$, as seen also in
Figure~\ref{fig3}. In Figure~\ref{fig4}b the behavior is very
similar: the error on the x-component is largest. In this case,
and since in Method IIa the cluster template is fixed, the scaling
is not as exactly $\propto N_{cl}^{1/2}$, reflecting the
inhomogeneities present in the cluster distribution. However,
these deviations are not very significant.

To study the effect of cluster inhomogeneities potentially present
in studies based on other catalogs, we carry a different analysis.
In Fig~\ref{fig4}c we excise clusters from the KAEEK catalog as a
function of galactic latitude. We plot $\langle
n_i^2\rangle^{-1/2}$ evaluated over the cluster distribution when
all clusters with $|b|\le b_{cut}$ are removed. Thick and thin
solid and dashed lines correspond to the $x,y,z$ components,
respectively. The dotted line is $\langle
n_i^2\rangle^{-1/2}=\sqrt{3}$, that would correspond to a cluster
catalog that samples the sky homogeneously. Since the mask removes
the data in the Galactic Plane, for $b_{cut}\la 20^\circ$ there is
little deviation from the KAEEK errors. For much larger values of
$b_{cut}$, the error on the $x$ and $y$ components increases while
the error on the $z$ component approaches that of the monopole.
Then, eq~(\ref{sigma_mon}) permits to write the error bars as
$\sigma_{(x,y,z)}=(1.12,1.05,0.87)\times15\sqrt{3/N_{\rm
cl}}\mu$K, i.e., the expected accuracy for each of the components
would be only 12 and 5\% worse that for an all sky survey,
compared with that of the monopole, while the $z$ component would
be 12\% better since the Galaxy removes the region of the sky
where there is no contribution to it.

Comparing the different panels in Figure~\ref{fig4}, we see that
$\sigma_y/\sigma_0$ may be smaller
than the value estimated from the geometry of the catalog as is
evident when comparing this ratio for $N_{cl}=1000$ in
Figure~\ref{fig4}b with Figure~\ref{fig4}c. However, while in (c)
the ratio of the errors is computed from the cluster geometry, in
(b) they are estimated from simulations whereby in Method IIa we
use simulations of the power spectrum of the filtered CMB data.
Since monopole and dipole are sensitive to different parity
multipoles (even vs odd), the slightly lower value of the dipole
components with respect to the monopole is reflecting a power
asymmetry between odd and even multipoles in the filtered map. So,
on average the monopole is larger than in a random sky and the
dipole is smaller. This effect introduces an extra variance and
enhances the differences between Fig~\ref{fig4}a and b.

When this paper was being completed, Keisler (2009) replicated the
analysis of KABKE1,2 compiling his own X-ray cluster catalog using
publicly available data. Analogously to KAEEK and this study, he
noticed that the errors on WMAP 3-year data were already dominated
by the residual CMB and not by the noise. He confirmed the
measured central dipole values of KABKE2, but claimed
significantly larger errors than KAEEK, particularly for the
$y$-component. (We note again that if the KABKE1,2 dipole
originated from primary CMB and/or noise, its magnitude should
display {\it no correlation} with the cluster luminosity threshold
that was demonstrated to exist in KAEEK). Specifically, in the
final configuration his catalog contained $\sim 700$ clusters and
his claimed errors were $\sigma^{\rm Keisler}_{x,y,z}\simeq
(1.7,1.7, 1.1) \mu$K. Those errors are larger than those quoted in
KAEEK. A small increment (of order of 10-15\%) can be accounted
for by his treatment of the errors using simulations of the CMB
sky around the theoretical $\Lambda$CDM model and thereby pumping
up the cosmic variance component (see Fig.~\ref{fig1}),  as well
as anisotropies in his catalog.
Keisler (2009) uses a catalog without recomputing cluster
properties from X-ray data, a procedure done in Kocevski \&
Ebeling (2006). That dataset is then less complete, especially at
low latitudes, but that in itself can account only for a small
increase in the errors. However, Keisler (2009) claims an increase
in errors by a factor of $>\sqrt{20}$ compared to KABKE2. Clearly,
the effect of residual CMB correlations between the $N_{DA}=8$
WMAP channels can at most increase the KABKE1,2 errors by a factor
of $\sqrt{N_{DA}}< \sqrt{8}$. (In reality, because the instrument
noise is also present, the errors on individual dipole components
in Table 2 of KABKE would be increased for 3-year WMAP data by a
factor of $\simeq \sqrt{6}$ to become $\la 1\mu$K at the largest
redshift bin.) A larger increase, as we have demonstrated above,
cannot happen. In our computations we do not reproduce Keisler
(2009) errors with proper analytical and numerical procedures,
even using his methodology.

\begin{figure}
\plotone{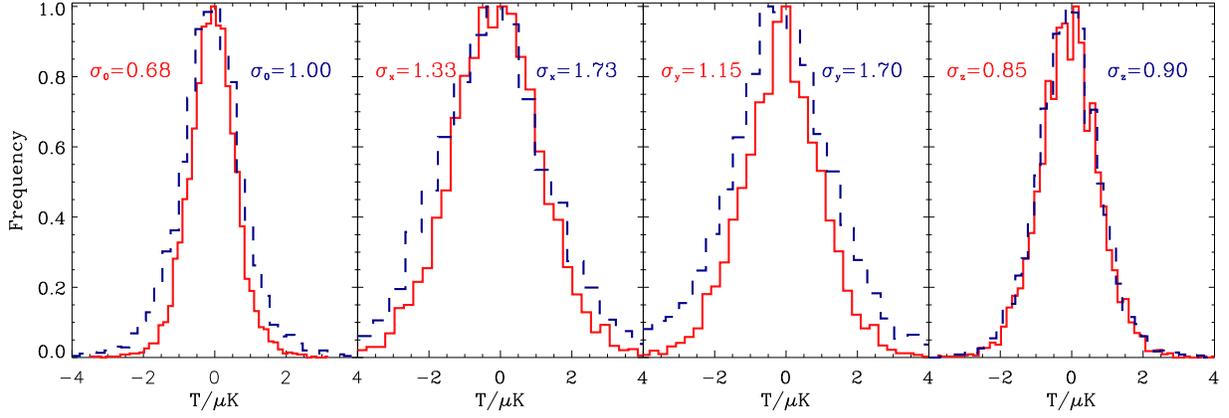} \caption{\small{Histograms  of 4,000
realizations of the CMB sky using Q1 DA parameters. Solid, dashed
lines correspond S1, S2 simulations; in S1 (S2) the monopole and
dipole outside the mask are (are not) removed. The left,
right rms dispersion corresponds to S1, S2, respectively. The figures
are given in micro Kelvin.
 }} \label{fig5}
\end{figure}

Interestingly, we recover the magnitude of his claimed errors if
one important aspect of the KABKE processing is omitted. When
working on simulated data such as in Method II, care must be taken
to replicate all the details of the data analysis done in
KABKE1,2. The filter must be constructed using the theoretical
model and the simulated data. Since only modes with $\ell\ge 4$
are used to generate the filtered map, by construction it has zero
monopole and dipole. But this map covers the full sky and is not
yet the correct model for the data. One needs to remove the
monopole and dipole outside the Galactic mask, as is done with the
filtered data in our processing. The full map has zero monopole
and dipole, but {\it the fraction of the sky outside the mask does
not}. To test this effect we carry out two sets of 4,000
simulations S1 and S2, starting with the same initial seed and
using Method II. In S1 simulations we removed monopole and dipole
{\it outside} the Galactic mask; in S2 we did not. In
Figure~\ref{fig5} we show the histograms with the distribution
(from left to right) of the monopole and the (x,y,z) components of
the dipole. The solid and dashed line shows the results for the S1
and S2 simulations, respectively. The labels on the left (right)
give the rms deviation for S1 (S2). The differences can be easily
explained: if the monopole and dipole are not subtracted the
measured monopole and dipole at cluster locations are not
different from zero simply because we are sampling the signal over
a very small fraction of the sky. Rather they are not zero because
we are measuring the monopole and dipole present on the fraction
of the sky outside the mask. We checked that when in the S1
simulations we add the variance of the monopole/dipole subtracted
outside the mask and the variance of the monopole/dipole computed
at cluster positions, we obtain exactly the variance measured in
the S2 simulations. For instance, the variances on
$(a_{0},a_{1x},a_{1y},a_{1z})$ outside the mask are
$(0.5,1.5,1.3,0.1)(\mu$K$)^2$. The variances on the
monopole/dipoles measured at the location of 1000 clusters of our
catalog are $(0.5,1.8,1.3,0.7)(\mu$K$)^2$; if added with the
previous variances, the monopole/dipoles error bar increase by
(46,36,41,5)\%, respectively. As expected we see that, since the
$z$ axis is perpendicular to the galactic plane, the error bars
are boosted preferentially in the $x$ and $y$ directions. This
explains that in S2 simulations the error in the monopole
$\sigma_0$ - as well as $\sigma_{x,y}$ - is larger than
$\sigma_z$. Only in the case of the faulty S2 processing do we
recover the magnitude of errors found by Keisler. We cannot claim
that this step was necessarily overlooked by him but we do find
this coincidence puzzling especially when considering the
deviation of his ratios of $\sigma_{1y}/\sigma_{1z}$ and
$\sigma_{1x}/\sigma_{1y}$ from the (analytically) explained ratios
(Fig. 4) and Fig. 1.

\section{The Statistical Significance of the ``Dark Flow''.}

Because of the correlations in the final filtered maps between the
eight WMAP DA's, the S/N of the Dark Flow measurement is smaller
than suggested in Table 2 of KABKE2 for individual dipole
components, although not fatally so. This was corrected in KAEEK,
where it was also demonstrated that the dipole correlates strongly
with the cluster X-ray luminosity $L_X$, as it should if the
dipole signal originated from the kSZ effect and not from the
primary CMB. The minimal S/N of the dark flow measurement is, of
course, given by the single DA map processing. Fig. 8 of KABKE2,
which plots the mean CMB temperature decrement over cluster pixels
versus the cosine of the angle between the cluster and the apex of
the motion, shows that KABKE1,2 already detect the dipole at
cluster positions at the $\simeq (2.5-3)\sigma$ level in {\it
each} of the eight DAs. The overall S/N cannot then be lower than
this floor level. KAEEK further increase this significance and
measure the motion to a much larger scale. The systematic
uncertainties in our calibration procedure do not yet allow us to
quantify the properties of the flow better, but we hope to
accomplish this task in the coming years.

In the KAEEK catalog, the error on the $y$ component is only 5\%
larger than what it would be for a homogeneous cluster catalog.
Future versions of the catalog will include clusters at higher
redshifts that will help to probe the velocity field on even
larger scales. A great effort is devoted to produce a spatially
homogeneous and flux limited sample. If the ``Dark Flow'' is but a
large scale flow that affects all the scales out to the horizon,
one could argue that the signal is uniform on the entire sky and
would be unaffected by anisotropies on the cluster distribution in
alternative catalogs, but this is not so, as Fig.~\ref{fig4}
indicates: incompleteness and asymmetries increase the error bars
and could make some cluster catalogs insensitive to the flow.

The original evidence in favor of the measurement being real were
three (KABKE1,2): {\it (a) the motion was found at cluster
positions, (b) it was persistent when the number of clusters
increased from $\la 150$ to $\ga 700$, (c) the dipole kSZ signal
was measured when the tSZ monopole vanished.} Since the thermal
and kinematic components are both generated by the X-ray gas, it
was thought that a measurement of the kSZ effect could be obtained
only when enough frequency coverage allowed to remove the thermal
contribution, because of their different frequency dependence.
However, in Atrio-Barandela et al (2008 - hereafter AKKE) we
showed - for the first time - that cluster gas distribution
follows an NFW profile (Navarro, Frenk \& White 1996). Then,
cluster temperature falls with radius and, by adding the
contribution from the cluster outskirts, the kinematic component
dominates over the thermal in the KAB method (KABKE2). If
clusters were isothermal, the thermal SZ signal dipole due to the
inhomogeneous distribution of the sky could be large enough to
make the kSZ effect undetectable at WMAP frequencies.

In KAEEK we provided further evidence in support of the cluster
bulk flow being a real effect. The cluster catalog used there was
large enough to allow the analysis to be carried out in luminosity
bins. The kSZ signal is $\Delta T_{ksz}\sim \tau v_{B}$. Since
$\tau$ is proportional to the cluster electron density, it
correlates with X-ray luminosity. If the velocity does not
correlate with cluster luminosity, for example if the flow is
homogeneous across the cluster sample, we would expect the dipole
evaluated at different cluster subsamples to be larger for the
more luminous clusters. In KAEEK we were able to carry out such
test by decomposing the sample in luminosity bins and the analysis
conclusively showed that {\it (d) the measured dipole correlates
with X-ray luminosity}, strengthening the evidence against a
possible undiagnosed systematic effect.

In KAEEK it was shown that clusters with the highest luminosity
dominate the S/N of the measured flow. To quantify the level of
statistical significance there, we generate 10,000 dipole
components drawn from a gaussian distribution with zero mean and
rms the measured error of each component as shown in Table 1 of
KAEEK. The significance is then the percentage of simulated values
that deviate from zero less than the measurement. For instance,
when we consider the measured dipoles for $L_X \geq 2\times
10^{44}$erg/sec clusters with $z\le 0.25$ we measured
$(a_{1x},a_{1y},a_{1z}) = (3.7\pm1.8, -4.1\pm 1.5, 4.1 \pm
1.5)\mu$K. If the dipoles $a_{1i}$ are Gaussian-distributed random
variables, the amplitude of the flow for these clusters is
detected at the 99.95\% level consistent with our simulations (in
Method I we find just 2 realizations out of 4,000 with such
parameters). For some other configurations in Table 1 of KAEEK the
confidence level would be even higher. Foreground contributions,
by their non-Gaussian nature, can in principle alter the above
percentiles, but the fact that our Universe lies so close to the
lines in Fig. 1 generated from pure primary CMB, implies that
foreground emissions contributions are small in our calculations.
We do not necessarily advocate the above levels to be highly
precise, but this discussion clearly shows that we recover a very
statistically significant dipole. While the dipole components are
less significant in lower $L_X$-bins, presumably because of the
lower $\tau$'s for these clusters, the $a_{1y}$ component is
always negative and $a_{1z}$ almost always positive in all three
$L_X$-bins, while the $a_{1x}$ component oscillates and is the
least accurately measured component. In this case, the possibility
that $a_{1x}$ is zero can be rejected at more than 95\% and
$a_{1z},a_{1z}$ at the 99\% confidence level. Due to the changing
sign, the measurements of the lower $L_X$ bins reduces the
significance of the detection of $a_{1x}$ and we can not claim any
measurement but the other two components are still significant at
more than 95\%. Finally, these probabilities would become even
higher if one folds in the directional coincidence of the
recovered dipole to that measured by Watkins et al (2009) from
galaxy surveys data on smaller scales, $\la 100$ Mpc.

\section{Possible, but negligible, $L_X$-dependent (SZ) systematics}

The only possible systematic effect that could mimic our
measurements would have to be present exclusively at cluster
positions, produce zero monopole and also give a dipole which
increases with increasing $L_X$. Such systematics cannot come from
primary CMB, and would have to originate from contributions by the
SZ components, which depend on $L_X$ in the appropriate manner.
Since in KAEEK we showed that the measured dipole correlates with
the X-ray luminosity threshold, it is important to discuss
possible $L_X$-dependent contributions even if only to rule them
out because of their negligible magnitudes. Given that we evaluate
the dipole at the aperture where the monopole vanishes, there are
three ways that could potentially confuse the measurement: 1)
Systematic effects that could fold the Doppler-shifting due to the
local motion into the tSZ contributions, 2) cross-talk effects
between the tSZ monopole and dipole terms in sparse/small samples
(Watkins \& Feldman 1995); and 3) inner motions of the
intracluster medium (ICM) as opposed to the coherent flow of the
entire cluster sample.

We discuss all three of these contributions below and demonstrate
that they are negligible. Before we go into the rest of the
section, we emphasize again that the dipole at cluster positions
is measured at zero monopole. That monopole vanishes within the
noise with 1-$\sigma$ uncertainty of $\simeq 15/\sqrt{N_{\rm
cl}}\mu$K or amplitudes significantly below $1\mu$K for $N_{\rm
cl}\ga 200$; the actual numbers are given in Table 1 of KAEEK.

The first two of these contributions come from the tSZ component,
while the latter would arise from the kSZ effect. Thermal and
kinematic SZ dipoles will differ in one very important aspect:
their frequency dependence. WMAP measures only in the
Rayleigh-Jeans part of the spectrum and for the Q, V and W bands
the change in amplitude is about $\la 30\%$, too small to be
distinguished (see KABKE2 for a discussion). A tSZ induced dipole
will change sign in the Wien part of the spectrum, while a kSZ
dipole will preserve it. The latter will be different from zero at
217GHz, the zero crossing frequency of the thermal component.
Although we show below that the tSZ induced contributions to the
dipole are very small, PLANCK with its large frequency coverage
covering both sides of 217 GHz will be definitive in this respect.

\subsection{Systematics due to tSZ shift from the local motion}

The intrinsic CMB dipole due to the motion of the Sun is over two
orders of magnitude larger than the measured cluster dipole. This
motion is known to be $u_\odot\simeq 370\pm 3$km/s in the
direction $(l,b)=(264^0,48^0)$, close to the direction
$(276^0,30^0)$ of the Local Group with respect to the same
reference frame (Kogut et al, 1993) and is not far within the
errors from the direction measured in KAEEK: $(290\pm 20, 30\pm
15)^0$. An undiagnosed systematic effect, present in the time
ordered data or in our pipeline that affect preferentially the tSZ
signal, could fold  the motion of the Sun into our measurement.
For example, a residual of the CMB all-sky dipole $(\Delta
T)_{res}$ coupled to the thermal SZ effect would correlate with
X-ray luminosity and would satisfy the same properties (a-d) as
the kSZ effect, except its frequency dependence. The amplitude of
such undiagnosed systematic dipole will be bound by $(\Delta
T)_{res}<(\Delta T)_{tSZ}(u_\odot/c)$. In AKKE we showed that the
tSZ amplitude of clusters in unfiltered maps is of the order of
$\sim -30\mu$K and this amplitude is reduced a factor of $\sim 3$
due to filtering (KABKE2). Then, any possible systematic effect
that correlates with cluster luminosity will be $(\Delta
T)_{res}<10^{-2}\mu$K, more than 2 orders of magnitude smaller
than the measured effect.

\subsection{Cross-talk from tSZ monopole in KAEEK sample}

Since clusters are not randomly distributed on the sky, the tSZ
signal will give rise to a non-trivial dipole signature that, in
principle, may confuse the kSZ dipole. The tSZ dipole for a random
cluster distribution is given by $a_{1m}^{\rm tSZ}\sim \langle
(\Delta T)_{\rm tSZ}\rangle \large({3/N_{\rm cl}}\large)^{-1/2}$
decreasing with increasing $N_{\rm cl}$. This decrease could be
altered if clusters are not distributed randomly and there may be
some cross-talk between the monopole and dipole terms especially
for small/sparse samples (Watkins \& Feldman 1995). As discussed
in KABKE2, the dipole from the tSZ component varies with the
cluster sub-sample, contrary to measurements, and also has
negligible amplitude because it is bound from above by the
remaining monopole amplitude of $\langle (\Delta T)_{\rm
tSZ}\rangle\ll 1\mu$K measured at the final aperture (see Table 1
of KAEEK).

In order to assess that there is no cross-talk between the
remaining monopole and dipole which may confuse the measured kSZ
dipole, we proceed in the same manner as in KABKE2 (see Fig 6
there) repeating the following experiment: 1) The tSZ and kSZ
components from the catalog clusters were modeled using cluster
parameters derived for our current catalog. To exaggerate the
effect of the cross-talk from the tSZ component, the latter was
normalized to $\langle (\Delta T)_{\rm tSZ}\rangle=-1\mu$K, a
value significantly larger than the monopoles in Table 1 of KAEEK
at which the final dipole was measured; the results for even
larger monopoles were also computed and can be scaled as described
below. For the kSZ component each cluster was given a bulk
velocity, $V_{\rm bulk}$, in the direction specified in Table 1 of
KAEEK, whose amplitude varied from 0 to 2,000 km/sec in 21
increments of 100 km/sec. The resultant CMB map was then filtered
and the CMB dipole, $a_{1m}({\rm cat})$, over the cluster pixels
computed for each value of $V_{\rm bulk}$. 2) At the second stage
we randomized cluster positions with $(l,b)$ uniformly distributed
on celestial sphere over the {\it full} sky for a net of 500
realizations for each value of $V_{\rm bulk}$. This random catalog
keeps the same cluster parameters, but the cluster distribution
now occupies the full sky (there is now no mask) and on average
does not have the same levels of anisotropy as the original
catalog. We then assigned each cluster the same bulk flow and
computed the resultant CMB dipole, $a_{1m}({\rm sim})$, for each
realization. The final $a_{1m}({\rm sim})$  were averaged and
their standard deviation evaluated.

\begin{figure}
\plotone{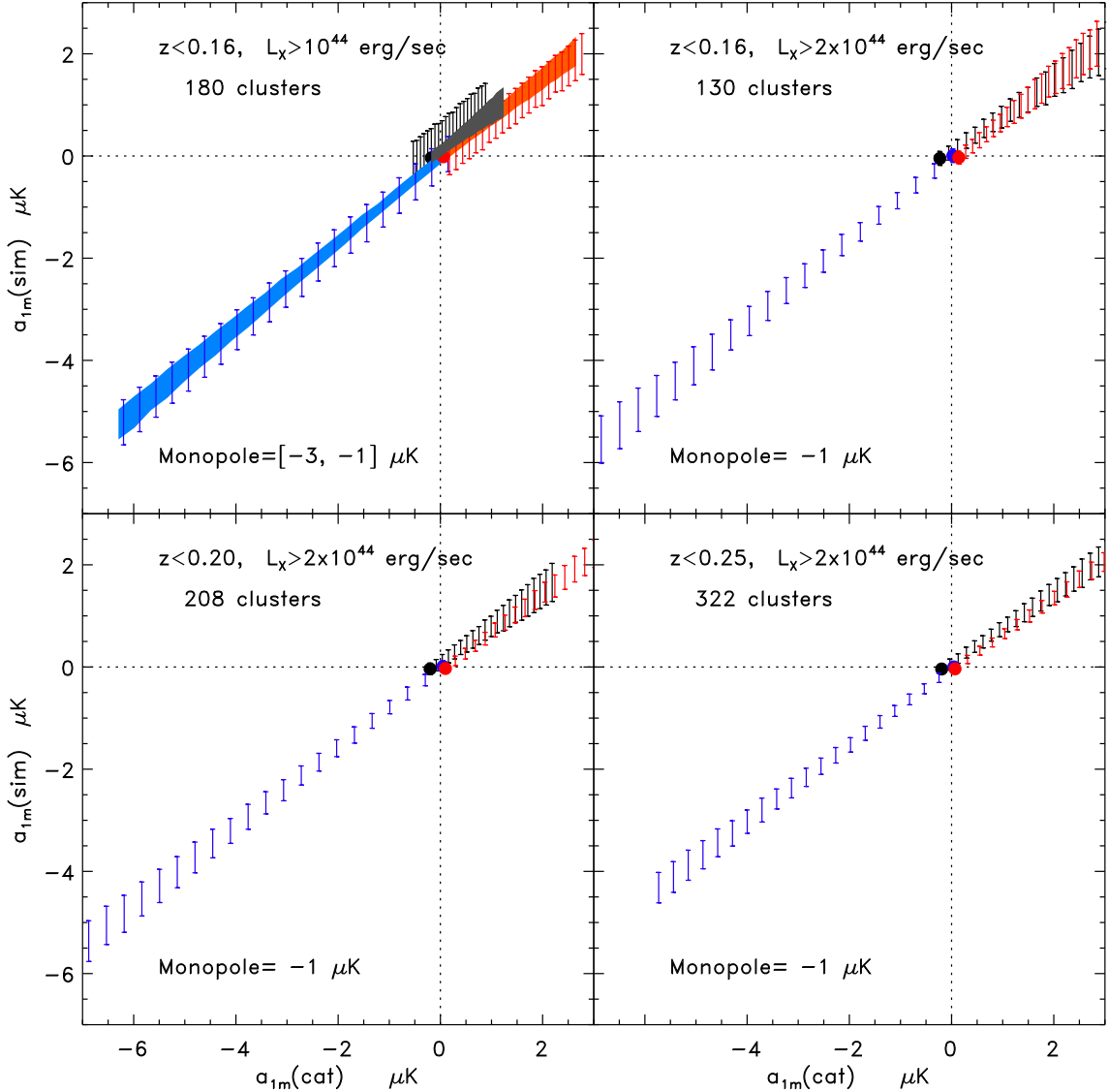} \caption{\small{The dipole coefficients for
simulated cluster distribution (random and, on average, isotropic)
are compared to that from the true catalog. (See text for
details). Each cluster in each catalog is given bulk flow of
$V_{\rm bulk}$ from 0 to 2,000 km/sec in increments of 100 km/sec
towards the apex of the motion from Table 1 of KAEEK. The results
from 500 simulated catalog realizations were averaged and their
standard deviation is shown in the vertical axis. Dotted lines
mark the zero dipole axis of the panels. The four most sparse
samples from Table 1 of KAEEK are shown which correspond to the
largest $L_X$-bins giving the best measured S/N. Black/blue/red
colors show the x/y/z components of $a_{1m}$. Filled circles of
the corresponding colors show the dipole components due to the
modeled tSZ component. The upper left panel shows the results for
two values of the monopoles: in the case of $\langle \Delta T_{\rm
tSZ}\rangle=-3 \mu$K the results are shown as individual error
bars; the case of $\langle \Delta T_{\rm tSZ}\rangle=-1 \mu$K is
shown with filled contours. All other panels show the results for
$\langle \Delta T_{\rm tSZ}\rangle=-1 \mu$K and our simulations
find good scaling with higher monopole values as described in the
text.
 }} \label{fig6}
\end{figure}

Fig. \ref{fig6} shows the comparison between the two dipoles for
each value of $V_{\rm bulk}$ for the most sparse sub-samples from
Table 1 of KAEEK. We also made the computations at tSZ monopole
values still larger than above (see upper left panel for one such
example). The overall contribution from the tSZ component to the
dipole is $\propto \langle \Delta T_{\rm tSZ}\rangle$, so in the
absence of cross-talk effects the amplitude of the scatter in the
simulated dipoles is made of two components: 1) remaining tSZ
$\propto \langle \Delta T_{\rm tSZ}\rangle$ and 2) genuine kSZ
dipole with amplitude $\propto V_{\rm bulk}$ to within the
calibration. One can see that there is no significant offset in
the CMB dipole produced by either the mask or the cluster true sky
distribution. The two sets of dipole coefficients are both
linearly proportional to $V_{\rm bulk}$ and to each other; in the
absence of any bulk motion we recover to a good accuracy the small
value of the tSZ dipole marked with filled circles. As discussed
in KABKE2, since the bulk flow motion is fixed in direction and
the cluster distribution is random, one expects the calibration
parameterized by $C_{1,100}$ to be different from one realization
to the next, e.g. in some realizations certain clusters may be
more heavily concentrated in a plane perpendicular to the bulk
flow motion and the measured $C_{1,100}$ would be smaller. In our
case, the mean $C_{1,100}$ differs by $\la 10\%$ suggesting that
our catalog cluster distribution is close to the mean cluster
distribution in the simulations. This difference in the overall
normalization would only affect our translation of the dipole in
$\mu$K into $V_{\rm bulk}$ in km/sec, but we note again the
systematic bias in the calibration resulting from our current
catalog modeling clusters as isothermal $\beta$-model systems
rather than the NFW profiles required by our observations (AKKE,
KABKE2). We have no progress to report on this issue beyond
discussion in sec. 8 of KABKE2 and this paper does not address the
measured velocity amplitude stemming from calibration; this work
is in progress and will be addressed after the recalibration of
our catalog has been successfully completed.

\subsection{Contribution from intracluster flows}

The intra-cluster medium (ICM) may not be at rest in the cluster
potential wells as a result of mergers during cluster formation
process. In principle, our measurement and interpretation then may
be affected by turbulent motions that give rise to a kSZ effect
that would be larger for the more massive clusters. However, since
the motions are randomly oriented with respect to the line of
sight, they will not produce a significant effect. In order, to
reach the value comparable to $V_{\rm bulk}\sim 1,000$ km/sec, a
typical cluster in our sub-sample of $N_{\rm cl}$ would need to
have thermal motions of $\sim V_{\rm bulk} N_{\rm cl}^{1/2}$, over
an order of magnitude larger than the velocity dispersion of
Coma-type clusters. Rather these motions will enter the overall
dispersion budget (noise, gravitational instability and this
component) around the coherent bulk flow component shown in Fig. 2
of KAEEK.

\section{Conclusions.}

We have analyzed the statistical significance of the results
presented in KAEEK. We have identified the main contributions to
the error budget: noise and the residual CMB contribution. While
the instrument noise was important in WMAP 1-year data, it was
much less so in the 3- and 5-year data. With our filtering scheme,
there remains a residual contribution due to cosmic variance,
which correlates at different frequencies and decreases only as
the number of clusters increases. We have discussed methods to
compute the errors and presented analytical discussion to estimate
the various contributions to the final error budget. Measuring
dipoles with a fixed template over simulated skies increases the
error bar in two respects: clusters do not sample the sky
homogeneously and different maps will have different CMB
residuals. Since the measured CMB sky in our Universe has less
power than the average $\Lambda$CDM realization, this also can
boost the errors, but by only $\sim$10-15\%. Also, the
inhomogeneities on the cluster distribution make the error on the
various dipole components different. For the $y$ component the
increment is about 5\% compared to the ideal case. We have argued
that a proper method to compute error bars would be to perform
random simulations of the measured power of the filtered maps
corresponding to the CMB sky of our Universe. However, we found
that the difference with taking random clusters outside the mask,
but using real data, was insignificant.

We have discussed the evidence supporting the existence of the
Dark Flow. Independently, different groups using galaxies as
tracers of the density or velocity field are showing the amplitude
and direction of the local flow that are consistent, albeit at a
much smaller scale, with the Dark Flow motion (Kocevski \& Ebeling
2006, Watkins et al 2009, Feldman et al and 2009, Lavaux et al
2009). This analysis with the forthcoming PLANCK data will provide
an important consistency check. With a scanning strategy different
from WMAP and with better frequency coverage, it will permit us to
characterize still better any possible undiagnosed systematic. The
217 GHz band with $\sim 5^\prime$ resolution will be specially
useful since it will allow to measure the kSZ signal from central
parts of the clusters in our catalog uncontaminated by the thermal
component.

This work was supported by NASA ADP grants NNG04G089G and
09-ADP09-0050. FAB acknowledges financial support from the Spanish
Ministerio de Educaci\'on y Ciencia (grant FIS2009-07238) and the
Junta de Castilla y Le\'on (grant GR-234).

\clearpage
\pagestyle{plain}

\clearpage

\clearpage

\clearpage

\clearpage

\clearpage

\end{document}